\documentclass{article}
\usepackage{graphicx}
\usepackage{amsmath,amssymb}
\usepackage{graphics,color,array,calc,rotating,epsfig,psfrag}
\numberwithin{equation}{section}
\usepackage{cite}
\usepackage{bm}
\usepackage{dcolumn}  
\usepackage{amsfonts}
\usepackage[mathscr]{eucal}
\def\be{\begin{equation}} \def\ee{\end{equation}}
\def\bea{\begin{eqnarray}} \def\eea{\end{eqnarray}}

\newcommand\prt{\partial}

\newcommand{\nn}{\nonumber}
\begin{document}
\baselineskip 18pt%
\begin{titlepage}
\vspace*{1mm}%
\hfill%
\vspace*{15mm}%
\hfill
\vbox{
    \halign{#\hfil         \cr
          } 
      }  
\vspace*{20mm}

\centerline{{\large {\bf Hamiltonian formalism of Minimal Massive Gravity}}}
\vspace*{5mm}
\begin{center}
{Davood Mahdavian Yekta \footnote{d.mahdavian@hsu.ac.ir}}\\
\vspace*{0.2cm}
{ Department of Physics, Hakim Sabzevari University, \\
P.O. Box 397, Sabzevar, Iran}\\
\vspace*{0.1cm}
\end{center}

\begin{abstract} 
In this paper, we study the three-dimensional minimal massive gravity (MMG) in the Hamiltonian formalism. At first, we define the canonical gauge generators as building blocks in this formalism and then derive the canonical expressions for the asymptotic conserved charges. The construction of a consistent asymptotic structure of MMG requires introducing suitable boundary conditions. In the second step, we show that the Poisson bracket algebra of the improved canonical gauge generators produces an asymptotic gauge group, which includes two separable versions of the Virasoro algebras. For instance, we study the BTZ black hole as a solution of the MMG field equations and the conserved charges give the energy and angular momentum of the Banados-Teitelboim-Zanelli (BTZ) black hole as a solution of the MMG field equations, and the conserved
charges give the energy and angular momentum of the BTZ black hole. Finally, we compute the black hole
entropy from the Cardy formula in the dual conformal field theory and show our result is consistent with the
value obtained by using the Smarr formula from the holographic principle.

\end{abstract} 

\end{titlepage}

\section{Introduction}
One of the technical obstacles to the quantization of general relativity (GR) is that the theory suffers from the lack of renormalization. Although a lot of attempts have been made to renormalize GR by adding higher curvature gravities, see for example\cite{Stelle:1976gc}, this work violates the unitarity of the theory which is a consistent condition in quantum gravity (the theory includes a massive spin-2 ghost mode). The 3D Einstein gravity is another choice to study quantum gravity \cite{Gott:1986bp}. The 3D gravity which is described by the Einstein-Hilbert(EH) action, with or without a cosmological constant \cite{Deser:1984tn}-\!\!\cite{Witten:1988hc}, has no physical local degrees of freedom. Adding a gravitational Chern-Simons (CS) term to the EH action, which is known as topologically massive gravity (TMG), gives a physical spin-2 mode \cite{Deser:1982vy}-\!\!\cite{Deser:1981wh}.  

Even though TMG might be a renormalizable and unitary theory of gravity in 3D bulk spacetime \cite{Deser:1990bj,Keszthelyi:1991ha}, it has also the bulk-boundary unitarity problem. From the gauge-gravity correspondence \cite{Maldacena:1997re}, when one considers the spin-2 mode of a unitary 3D bulk gravity on an asymptotically anti-de Sitter ($AdS$) background, there might be a unitary conformal field theory (CFT) on the 2D boundary. The unitarity of the dual CFT implies the existence of positive central charges in the symmetry group algebra. On the other hand, to have a positive energy Banados-Teitelboim-Zanelli (BTZ) black hole \cite{Banados:1992wn} in TMG, Newton's constant must be positive, which gives a negative central charge in the dual boundary theory\cite{Li:2008dq}. 

It seems that the above problem will be resolved if we consider the bulk theory at a critical point (known as chiral gravity\cite{Li:2008dq,Maloney:2009ck}), at which the negative central charge vanishes. Specifically, at this point, the spin-2 mode is replaced by a logarithmic mode, which results in a nonunitary logarithmic CFT at the dual boundary theory \cite{Maloney:2009ck,Grumiller:2008es}. However, other modified 3D theories have been introduced such as new massive gravity (NMG) and its extensions \cite{Bergshoeff:2009hq}-\!\!\cite{Ghodsi:2010ev},  which suffer from a similar problem in critical points \cite{Liu:2009kc}.

Recently, a new version of 3D massive gravity has been proposed in Ref. \cite{Bergshoeff:2014pca}, which includes one massive degree of freedom similar to TMG and is called the minimal massive gravity (MMG). Not only this theory is unitary in the bulk, but it also has a unitary dual CFT on the boundary, of course for some values in the parameter space of the theory.  A particular feature of this theory is that it has no Lagrangian in the metric formalism while is formulated in the first-order canonical structure. Some aspects of this theory such as black hole solutions and the existence of the chiral points have been considered in Refs. \cite{Arvanitakis:2014yja}-\!\!\cite{Tekin:2014jna}.

Here we study MMG in the canonical Hamiltonian first-order formalism. We will find expressions for the gauge generators from a constrained Hamiltonian theory \cite{Castellani:1981us}-\!\!\cite{Blagojevic:2010jv}. The asymptotic conserved charges can be constructed from these generators, which in the case of BTZ black hole solution give the energy and angular momentum. We also discuss about the symmetries of the asymptotic configuration of the theory and show that the analysis of the asymptotic symmetry algebra for the BTZ solution includes two Virasoro algebras with two different central charges in the left and right sectors. 

Hawking and Bekenstein have shown in Refs. \cite{Bardeen:1973gs} and\cite{Bekenstein:1973ur} that the black holes can provide us thermodynamical systems in GR. So we will discuss the thermodynamics of the BTZ black hole, especially its entropy. Since MMG is constructed from a field equation in the metric formalism instead of a Lagrangian, we can not compute the entropy from the Wald formula \cite{Wald:1993nt}. On the other side, we can obtain it from the Smarr formula \cite{Smarr:1972kt} and then verify that our result is in accordance with the one obtained from the Cardy formula \cite{Cardy:1986ie}.

This paper is organized as follows: In section 2, we briefly discuss the MMG theory and derive the field equations of motion. In section 3, the canonical structure of the theory is constructed, and the general form of the gauge generators is calculated. In section 4, we clarify the application of this formalism by studying the BTZ black hole as a solution to the MMG field equations, and then we investigate the asymptotic behavior of the theory. In Section 5, we give a summary of the discussions and concluding results.

The metric in the local Lorentz frame is mostly minus $\eta_{ij}=(+,-,-)$; the Latin indices $(i,j,k,\dots)$ and the Greek indices $(\mu,\nu,\lambda,\dots)$ refer, respectively, to the local Lorentz and coordinate frames and run over $0,1,2$, while the letters $(\alpha,\beta,\gamma,\dots)$ run over $1,2$; both Levi-Civit\'{a} antisymmetric tensors $\varepsilon^{ijk}$ and $\varepsilon^{\mu\nu\rho}$ are normalized as $\varepsilon^{012}=1$. The space-time metric is a bilinear combination of the triad fields: 
\bea \label{metric} &g=g_{\mu\nu}\, dx^{\mu}\otimes dx^{\nu}=\eta_{ij}\, e^{i}\otimes e^{j}\,,&\nn\\
&g_{\mu\nu} =\eta_{ij}{e^{i}}_{\mu} {e^{j}}_{\nu}\,.&
\eea
\section{Minimal Massive Gravity}
The linearization of TMG around a maximally symmetric background gives a negative energy massive graviton which threats the unitarity of the theory \cite{Li:2008dq}. This problem can be resolved if we change the sign of the Newton's constant which due to this fact, the central charge of the dual boundary CFT becomes negative. A minimal construction of TMG has been introduced in \cite{Bergshoeff:2014pca} which can resolve this problem. 

The Einstein field equation of TMG \cite{Deser:1981wh} is given by
\be \label{TMG} \frac{1}{\mu}\, C_{\mu\nu}+\sigma\, G_{\mu\nu}+\Lambda_{0}\, g_{\mu\nu}=0\,,\ee
where $G_{\mu\nu}$ and $C_{\mu\nu}$ are the Einstein and symmetric traceless Cotton tensors and $\sigma,\mu,\Lambda_0$ are constants with mass dimensions $0,1$ and $2$, respectively. In Ref. \cite{Bergshoeff:2014pca}, a new field equation is defined for TMG as a minimal construction, 
\be \label{MMG} \frac{1}{\mu}\, C_{\mu\nu}+\bar\sigma\, G_{\mu\nu}+\bar\Lambda_{0}\, g_{\mu\nu}=-\frac{\gamma}{\mu^2}\,J_{\mu\nu},\ee 
with the symmetric tensor defined as
\be J_{\mu\nu}\equiv\frac{1}{2 |g|}\, {\varepsilon_{\mu}}^{\rho\lambda}\,{\varepsilon_{\nu}}^{\tau\eta} S_{\rho\tau} S_{\lambda\eta}\,,\ee
where $S_{\mu\nu}$ is the Schouten tensor,
\be \label{CST} C_{\mu\nu}=\frac{1}{\sqrt{|g|}}\, {\varepsilon_{\mu}}^{\tau\eta} D_{\tau} S_{\eta\nu}\,\,,\qquad S_{\mu\nu}=R_{\mu\nu}-\frac14 g_{\mu\nu} R\,. \ee
In Eq. (\ref{MMG}), $\gamma$ is a non-zero dimensionless constant, and $\sigma$, $\Lambda_{0}$ in the case of TMG (\ref{TMG}) are replaced by $\bar\sigma$, $\bar\Lambda_{0}$ for MMG, since it is not obvious they should be remain equal to the initial values. Because the linearization of (\ref{MMG}) around a maximally symmetric background gives the linearized TMG with modified coefficients \cite{Tekin:2014jna}, MMG has the same degrees of freedom as TMG. 

As we see, this theory is constructed in terms of a field equation (\ref{MMG}) which is not obtained from a metric action similar to TMG and NMG. So for further studies, it would be better to consider this theory in another formalism. The $3$-form Lagrangian of MMG in the Chern-Simons-like formalism is given by 
\be \label{LMMG1} L_{MMG}=-\sigma\,e\cdot R+\,\frac{\Lambda_0}{6}\, e\cdot e\times e+h \cdot T+\frac{1}{2\mu}\left(\omega \cdot d \omega+\frac13\, \omega\cdot \omega\times\omega\right)+\frac{\alpha}{2}\,e\cdot h\times h\,,\ee
where as before $\sigma\,,\alpha$ are dimensionless parameters, $\Lambda_0$ is a cosmological constant term with mass-squared dimension, and $\mu$ is a parameter of dimension 1. It has been shown that MMG resolves the problem of unitarity (no ghost/no tachyon) for some values of these parameters\cite{Bergshoeff:2014pca}. 

To have a torsion free condition, we use a Lagrange multiplier field $h$, which, similar to spin connection $\omega$, is an odd parity one-dimensional field. For subsequent aims, we rewrite the Lagrangian (\ref{LMMG1}) in some useful form, 
\bea \label{LMMG2} L_{MMG}=\frac12\, \varepsilon^{\mu\nu\rho}\Big[\!\!\!\!&-&\!\!\!\!\sigma\,{e^{i}}_{\mu} R_{i\,\nu\rho}+\frac{\Lambda_0}{3}\,{\varepsilon_{ijk}}\, {e^{i}}_{\mu} {e^{j}}_{\nu} {e^{k}}_{\rho}+{h^{i}}_{\mu}  T_{i\,\nu\rho}\nn\\&+&\!\!\!\frac{1}{\mu}\left({\omega^{i}}_{\mu}\prt_{\nu} {\omega_{i}}_{\rho}+\frac13\,{\varepsilon_{ijk}} {\omega^{i}}_{\mu} {\omega^{j}}_{\nu} {\omega^{k}}_{\rho} \right) +\alpha\,{\varepsilon_{ijk}}\, {e^{i}}_{\mu} {h^{j}}_{\nu} {h^{k}}_{\rho}\Big]\,,\eea
where the fields ${e^{i}}_{\mu}, {\omega^{i}}_{\mu}, {h^{i}}_{\mu}$ are three Lorentz 1-forms. The field strengths $T^{i}(\omega)$ and $R^{i}(\omega)$ are Lorentz covariant torsion and curvature $2$-forms given by
\be T^{i}=de^{i}+{\varepsilon^{i}}_{jk}\,\omega^{j} e^{k}\,,\qquad R^{i}=d\omega^{i}+\frac12 {\varepsilon^{i}}_{jk}\,\omega^{j}\omega^{k}\,.\ee

In the case of $\alpha=0$, the Lagrangian (\ref{LMMG1}) gives rise to cosmological topological massive gravity in the torsion-free limit \cite{Deser:1984tn}. In addition, for $\mu \to \infty$ and $\sigma= 1$, it reduces to the usual 3D cosmological gravity with constant curvature \cite{Deser:1984dr}. Taking variations of the Lagrangian (\ref{LMMG2}) with respect to the $1$-form fields ${e^{i}}_{\mu}, {\omega^{i}}_{\mu}$ and ${h^{i}}_{\mu}$ yields the field equations (in addition to some total derivative terms)
\bea \label{EOM}
0\!\!\!\!&=&\!\!\!\!\varepsilon^{\mu\nu\rho}\left(-\sigma\,R_{i\nu\rho}+\Lambda_0\,{\varepsilon_{ijk}} {e^{j}}_{\nu} {e^{k}}_{\rho}+D_{\nu} {h_{i}}_{\rho}+\alpha\,{\varepsilon_{ijk}} {h^{j}}_{\nu} {h^{k}}_{\rho}\right)\,,\nn\\
0\!\!\!\!&=&\!\!\!\!\varepsilon^{\mu\nu\rho}\left(-\sigma\, T_{i\nu\rho}+{\varepsilon_{ijk}}\, {e^{j}}_{\nu} {h^{k}}_{\rho}+{\mu}^{-1}\, R_{i\nu\rho}\right)\,,\\
0\!\!\!\!&=&\!\!\!\!\varepsilon^{\mu\nu\rho}\left(T_{i\nu\rho}+2\,\alpha\, {\varepsilon_{ijk}}\, {e^{j}}_{\nu} {h^{k}}_{\rho}\right)\,,\nn
\eea
where $``\,D(\omega)=d+\omega\times\,"$ is the Lorentz covariant derivative. One can rewrite the equations of (\ref{EOM}) as
\bea \label{feqs}
0\!\!\!\!&=&\!\!\!\!D(\Omega) h-\frac{\alpha}{2} h\times h+\sigma\mu(1+\sigma \alpha) e\times h+\frac{\Lambda_0}{2} e\times e\,,\nn\\
0\!\!\!\!&=&\!\!\!\!R(\Omega)+\frac{\alpha\Lambda_0}{2} e\times e+\mu(1+\sigma\alpha)^2 e\times h\,,\\
0\!\!\!\!&=&\!\!\!\! T(\Omega)\,,\nn
\eea
where the last line represents the torsion-free condition for connection field $\Omega=\omega+\alpha\,h$, i.e.,
\be \label{tfc} T^{i}=de^{i}+{\varepsilon^{i}}_{jk}\,\Omega^{j} e^{k}=0\,.\ee 

To have the same local degrees of freedom as TMG, we must have $(1+\sigma \alpha)\neq 0$ in parameter space, While for equal condition $(1+\sigma \alpha)= 0$, the theory becomes a CS-like theory with no degrees of freedom \cite{Witten:1988hc}. The second equation of (\ref{feqs}) yields
\be \label{hf} h_{\mu\nu}=-\frac{1}{\mu (1+\sigma\alpha)^2}\left[ S_{\mu\nu}+\frac{\alpha\Lambda_0}{2} g_{\mu\nu}\right]\,,\ee
where by substituting (\ref{hf}) in the first equation of (\ref{feqs}) and comparing the resultant equation by the metric field equation (\ref{MMG}), we conclude that
\be \gamma=-\frac{\alpha}{(1+\sigma\alpha)^2}\,,\quad\bar\Lambda_0=\Lambda_0\left[1+\sigma\alpha-\frac{\alpha^3\Lambda_0}{4\mu^2(1+\sigma\alpha)^2}\right]\,,\quad \bar\sigma=\sigma+\alpha\left[1+\frac{\alpha\Lambda_0}{2\mu^2(1+\sigma\alpha)^2}\right]\,.\ee
In other words, we have a canonical action (\ref{LMMG1}) which gives the field equation (\ref{MMG}).
\section{Canonical structure} \label{sec3}
In this section, we will consider the structure of the MMG as a gauge theory described by (\ref{LMMG2}) in the canonical formalism \cite{Castellani:1981us}. The gauge symmetries determine the physical content of any gauge theory by means of some gauge generators which lead to a number of conserved charges. Often there are two classes of conserved charges: the exact ones associated to the symmetries of the background solution and the asymptotic ones related to the symmetries close to the infinity or at the boundary. The calculation of these charges and their properties in the canonical formalism is the underling idea of our work hereafter. 

Here, we will not discuss this procedure in detail while enumerating the substantial constructions to obtain the gauge generators of the MMG theory as follows \cite{Castellani:1981us},\cite{Blagojevic:2008bn} :
\begin{enumerate}
	\item The canonical momenta $({\pi_{i}}^{\mu}, {\Pi_{i}}^{\mu}, {p_{i}}^{\mu})$ are, respectively, defined for the Lagrangian variables \\$({e^{i}}_{\mu}, {\omega^{i}}_{\mu}, {h^{i}}_{\mu})$,
	\be \label{cms} {\pi_{i}}^{\mu}\equiv\frac{\prt L}{\prt {\dot {e}^{i}}_{\mu}}\,,\qquad {\Pi_{i}}^{\mu}\equiv\frac{\prt L}{\prt {\dot {\omega}^{i}}_{\mu}}\,,\qquad {p_{i}}^{\mu}\equiv\frac{\prt L}{\prt {\dot {h}^{i}}_{\mu}}\,, \ee
	where, by using the Lagrangian (\ref{LMMG2}), we can define the following set of primary constraints: 
	\bea \label{pc} &&{\phi_{i}}^{0}\equiv {\pi_{i}}^{0} \approx 0\,,\qquad {\phi_{i}}^{\alpha}\equiv {\pi_{i}}^{\alpha}-\varepsilon^{0\alpha\beta} h_{i\beta} \approx 0\,,\nn \\
	&&{\Phi_{i}}^{0}\equiv {\Pi_{i}}^{0} \approx 0\,,\qquad {\Phi_{i}}^{\alpha}\equiv {\Pi_{i}}^{\alpha}+\varepsilon^{0\alpha\beta} (\sigma e_{i\beta}-{\mu}^{-1}\, \omega_{i\beta}) \approx 0\,,\nn\\
	&& {\psi_{i}}^{\mu}\equiv {p_{i}}^{\mu}\,\approx 0\,.
	 \eea
	 The canonical Hamiltonian in this construction is expressed as
	 \bea
	&& \mathcal{H}_{c}={e^{i}}_{0} \mathcal H_{i}+{\omega^{i}}_{0} \mathcal K_{i}+{h^{i}}_{0} \mathcal T_{i}+\prt_{\alpha} \mathcal S^{\alpha}\,,\\
&&	\mathcal H_{i}=- \varepsilon^{0 \alpha\beta} \left(-\sigma R_{i\alpha\beta}+\Lambda_0\, \varepsilon_{ijk}\, {e^{j}}_{\alpha} {e^{k}}_{\beta}+\alpha\, \varepsilon_{ijk} {h^{j}}_{\alpha} {h^{k}}_{\beta}+D_{\alpha} h_{i\beta}\right)\,,\nn\\
&&	\mathcal K_{i}=- \varepsilon^{0 \alpha\beta} \left(-\sigma T_{i\alpha\beta}+\varepsilon_{ijk}\, {e^{j}}_{\alpha} {h^{k}}_{\beta}+{\mu}^{-1} R_{i\alpha\beta}\right)\,,\nn\\
&&	\mathcal T_{i}=- \varepsilon^{0 \alpha\beta} \left( T_{i\alpha\beta}+2\, \varepsilon_{ijk} \,{e^{j}}_{\alpha} {h^{k}}_{\beta}\right)\,,\nn\\
&&	\mathcal S^{\alpha}=\varepsilon^{0 \alpha\beta} \left({\omega^{i}}_{0}[-\sigma {e_{i\beta}}+{\mu}^{-1} \omega_{i\beta}]+{e^{i}}_{0}{h_{i\beta}}\right)\,,\nn
	 \eea
	 where the last term in $\mathcal H_{c}$ is the surface term of (\ref{LMMG2}). The total Hamiltonian constructed from the primary constraints is given by  
	 \be \mathcal{H}_{T}={e^{i}}_{0} \mathcal H_{i}+{\omega^{i}}_{0} \mathcal K_{i}+{h^{i}}_{0} \mathcal T_{i}+{u^{i}}_{\mu} {\phi^{\mu}}_{i}+{v^{i}}_{\mu} {\Phi^{\mu}}_{i}+{w^{i}}_{\mu} {\psi^{\mu}}_{i}+\prt_{\alpha} \mathcal S^{\alpha}\,,\ee
	 and the consistency conditions can be written as the approach introduced in Refs. \cite{Castellani:1981us},\!\!\cite{Blagojevic:2008bn}. These conditions guarantee the primary constraints ${\pi_{i}}^{0}$, ${\Pi_{i}}^{0}$ and ${p_{i}}^{0}$ yield the secondary constraints,
	 \be\label{sc} \mathcal H_{i}\approx0\,,\quad \mathcal K_{i}\approx0\,,\quad \mathcal T_{i}\approx0\,, \ee
	 and due to the remaining primary constraints ${\phi_{i}}^{\alpha}$, ${\Phi_{i}}^{\alpha}$, and ${p_{i}}^{\alpha}$, the multipliers ${u^{i}}_{\mu}$, ${v^{i}}_{\mu}$, and ${w^{i}}_{\mu}$ are determined. Some of these consistency conditions are derived in the Appendix.
	 \item The canonical structure of the asymptotic symmetry is described by the canonical gauge generators
	 \bea \label{GG}
	 G\!\!\!\!&=&\!\!\!-G_1-G_2\,,\\
	 G_{1}\!\!\!\!&=&\!\!\! \dot\xi^{\rho}\left({e^{i}}_{\rho}{\pi_{i}}^{0}+{\omega^{i}}_{\rho}{\Pi_{i}}^{0}+{h^{i}}_{\rho}{p_{i}}^{0}\right)\nn\\
	 &&+\,\xi^{\rho}\left[ {e^{i}}_{\rho} \bar{\mathcal {H}}_{i}+{\omega^{i}}_{\rho} \bar{\mathcal {K}}_{i}+{h^{i}}_{\rho} \bar{\mathcal {T}}_{i}+(\prt_{\rho} {e^{i}}_{0}) {\pi_{i}}^{0}+(\prt_{\rho} {\omega^{i}}_{0}) {\Pi_{i}}^{0}+(\prt_{\rho} {h^{i}}_{0}) {p_{i}}^{0}\right]\,,\nn\\
	 G_2\!\!\!\!&=&\!\!\! \dot{\theta}^{i}{\Pi_{i}}^{0}+\theta^{i}\left[\bar{\mathcal K}_{i}-\varepsilon_{ijk}\left({e^{j}}_{0}{\pi}^{k0}+{\omega^{j}}_{0}{\Pi}^{k0}+{h^{j}}_{0}{p}^{k0}\right)\right]\,.\nn
	 \eea
	 where the dot sign is the time derivative and the factor of $\frac{1}{8\pi G} \int d^3x$ is omitted for simplicity. The local Poincar\'{e} gauge transformations (PGT) are 
	 \bea \label{PGT}
	 \delta_{0}{e^{i}_{\mu}}\!\!\!&=&\!\!\!-{\varepsilon^{i}}_{jk} {e^{j}}_{\mu} \theta^{k}-(\prt_{\mu} \xi^{\rho}){e^{i}}_{\rho}-\xi^{\rho} \prt_{\rho}{e^{i}}_{\mu}\,,\nn\\	 \delta_{0}{\omega^{i}_{\mu}}\!\!\!&=&\!\!\!-\nabla_{\mu}\theta^{i}-(\prt_{\mu} \xi^{\rho}){\omega^{i}}_{\rho}-\xi^{\rho} \prt_{\rho}{\omega^{i}}_{\mu}\,,\\
	 \delta_{0}{h^{i}_{\mu}}\!\!\!&=&\!\!\!-{\varepsilon^{i}}_{jk} {h^{j}}_{\mu} \theta^{k}-(\prt_{\mu} \xi^{\rho}){h^{i}}_{\rho}-\xi^{\rho} \prt_{\rho}{h^{i}}_{\mu}\,.\nn
\eea  
The symmetric parts of these equations give the asymptotic local translations $\xi^{\mu}$, and the antisymmetric ones give the local rotations $\theta^{i}$ of Poincar\'{e} transformations \cite{Blagojevic:2008bn}. For example, multiplying the first of (\ref{PGT}) by ${e_{i\nu}}$ and using (\ref{metric}) yields the general transformation of the metric
\be \label{gt} \delta_{0}G_{\mu\nu}= -(\prt_{\mu} \xi^{\rho})g_{\nu\rho}-(\prt_{\nu} \xi^{\rho})g_{\mu\rho}-\xi^{\rho} \prt_{\rho}g_{\mu\nu}\,.\ee 
 \item The variation of $G_2$ produces a total derivative term which, by choosing the consistent boundary conditions, given in the next section, has vanishing contribution after integration. In the next step, looking to the variation of $G_1$ and after some substitutions, we have
\bea \label{delg1}
\delta G_1\!\!\!\!&=&\!\!\!\!\ \xi^{\rho} \left( {e^{i}}_{\rho} \,\delta{\mathcal H_{i}}+{\omega^{i}}_{\rho} \,\delta{\mathcal K_{i}}+{h^{i}}_{\rho} \,\delta{\mathcal T_{i}}\,\right)+\prt \mathcal O+R\\
\!\!\!\!&=&\!\!\!\!2 \,\varepsilon^{0\alpha\beta}\,\xi^{\rho}\prt_{\alpha}\left[{e^{i}}_{\rho}\,(\sigma\, \delta \omega_{i\beta}-\frac12\, \delta h_{i\beta})+{\omega^{i}}_{\rho}\,(\sigma \,\delta e_{i\beta}-\,\frac{1}{\mu}\,\delta \omega_{i\beta})-{h^{i}}_{\rho}\, \delta e_{i\beta}\right]+\prt \mathcal O_2+R\,,\nn
\eea 
where the term $\prt \mathcal O_2$ is a boundary term that vanishes after integration and $R$ includes some regular terms. Hereafter, when we use $\mathcal O_{n}$, it has the distance behavior as $\sim r^{-n}$, so using the Stokes theorem
\be \label{ST} \int_{\mathcal M_2} d^2 x\prt_{\alpha}v^{\alpha}=\int_{\prt \mathcal M_2} v^{\alpha} df_{\alpha}=\int_{0}^{2\pi} v^1 d\varphi\qquad (df_{\alpha}=\varepsilon_{\alpha\beta}dx^{\beta})\,,\ee
 the second term in (\ref{delg1}) has no contribution to the asymptotic conserved charges. Here, the boundary of ${\mathcal M_2}$ is a circle at infinity parametrized by the angular coordinate $\varphi$.

\item We can write the relation (\ref{delg1}) as 
\be
\delta G_1=\prt_{\alpha} (\xi^{0} \delta \mathcal E^{\alpha}+\xi^{2}\delta \mathcal M^{\alpha})\,,
\ee
where 
\bea \label{eam}
\mathcal E^{\alpha}\!\!\!&=&\!\!\!2\, \varepsilon^{0\alpha\beta}\,\left[{e^{i}}_{0}\,(\sigma\, \delta \omega_{i\beta}-\frac12\, \delta h_{i\beta})+{\omega^{i}}_{0}\,(\sigma \,\delta e_{i\beta}-\,\frac{1}{\mu}\,\delta \omega_{i\beta})-{h^{i}}_{0}\, \delta e_{i\beta}\right]\,,\nn\\
\mathcal M^{\alpha}\!\!\!&=&\!\!\!2\,\varepsilon^{0\alpha\beta}\left[{e^{i}}_{2}\,(\sigma\, \delta \omega_{i\beta}-\frac12\, \delta h_{i\beta})+{\omega^{i}}_{2}\,(\sigma \,\delta e_{i\beta}-\,\frac{1}{\mu}\,\delta \omega_{i\beta})-{h^{i}}_{2}\, \delta e_{i\beta}\right]\,.
\eea

We look for the conserved charges which are related to the energy and angular momentum of the system and are obtained by choosing the diffeomorphisms as $\xi^{0}=1$ and $\xi^{2}=1$, such that
\be \label{eam2} E=\int_{0}^{2\pi}   \mathcal E^1\,d\varphi\quad,\quad J=\int_{0}^{2\pi}   \mathcal M^1\,d\varphi\,.\ee
\end{enumerate}

\section{Black Holes in MMG}
Any solution of the 3D Einstein gravity with a negative cosmological constant is automatically a solution of MMG. So the BTZ solution constitutes a candidate for the black hole object of the theory as $AdS_3$ as a vacuum solution.
There is a parameter space for constant values ($\sigma,\alpha,\mu,C,\Lambda_0$) in this theory which encourages us to exemplify the physical quantities for the BTZ black holes. 

\subsection{Canonical BTZ solution}
The most well-known maximally symmetric black hole solution for 3D gravity is the BTZ black hole for which its line element of space-time in the Arnowitt-Deser-Misner
(ADM) \cite{Arnowitt:1962hi} form is given by 
\be \label{btz1} ds^2=N^2 dt^2-N^{-2} dr^2-r^2(d\varphi+N_{\varphi} dt)^2\,,\ee
where the functions $N, N_{\varphi}$ are
\be N^2=\frac{(r^2-r_{+}^2)(r^2-r_{-}^2)}{r^2 l^2}\,,\quad N_{\varphi}=\frac{r_{+} r_{-}}{l\, r^2}\,.\ee
and $r_{+}$ , $r_{-}$ are the outer and inner horizons of BTZ black hole, respectively. Note that we use the mostly minus signature for the metric of space-time to be consistent by our convention for $\eta_{ij}$ in the canonical formalism. This is a maximally symmetric solution, such that
\be \label{mss} R^{i}=\frac{\Lambda}{2}\,{\varepsilon^{i}}_{jk} e^{j} e^{k}\,\ee
and the relation between cosmological parameter and $l$ is $\Lambda=-{1}/{l^2}$. The components of triad 1-form $e^{i}$ in the first order formalism for (\ref{btz1}) are 
\be \label{btz2} e^{0}=N dt\,,\quad e^{1}=N^{-1} dr\,,\quad e^{2}=r(d\varphi+N_{\varphi} dt)\,,\ee
and the components of spin connection $\Omega^{i}$ are computed from the torsion-free condition (\ref{tfc}) as
\be \Omega^{0}=-N d\varphi\,,\quad \Omega^{1}=N^{-1} N_{\varphi} dr\,,\quad \Omega^{2}=-\frac{r}{l^2}\, dt-rN_{\varphi} d\varphi\,.\ee

Substituting the relation (\ref{mss}) in the second equation of (\ref{feqs}) gives 
\be \label{he} h^{i}=\mu C e^{i}\,,\ee
 where the constant $C$ is defined in terms of parameter space,
\be \label{ps} C\equiv-\,\frac{(\Lambda+\alpha \Lambda_0)}{2\mu^2 (1+\sigma\alpha)^{2}}\,.\ee 
\subsection{Conserved charges}
As mentioned in section {\ref{sec3}}, there are two types of conservation laws according to asymptotic symmetries. In fact, those transformations that allow the field configurations under consideration to remain asymptotically invariant are asymptotic symmetries. They generate the known asymptotic symmetry group algebra \cite{BH}. On the other hand, not only should the asymptotic symmetries be invariant under the isometry group of asymptotically $AdS_3$ backgrounds, $SL(2,R)\times SL(2,R)$, but they should also have well-defined canonical generators. There is no doubt that we must use some suitable boundary conditions that respect the above conditions .

Thus, in the case of BTZ solution, the asymptotic behavior of triads are given by 
\be \label{bc1}
{e^{i}}_{\mu}={e^{i}}_{\mu}+{E^{i}}_{\mu}\,,\quad {e^{i}}_{\mu}=\left(\begin{array}{ccc}
 r/l & 0 & 0 \\ 
0 & l/r& 0 \\ 
0 & 0 & r 
\end{array} \right)\,,\quad
{E^{i}}_{\mu}\sim\left(\begin{array}{ccc}
 \mathcal O_{1} & \mathcal O_{4} & \mathcal O_{1} \\ 
\mathcal O_{2} & \mathcal O_{3} & \mathcal O_{2} \\ 
\mathcal O_{1} & \mathcal O_{4} & \mathcal O_{1} 
\end{array} \right)\,,
\ee
and for the connection field
\be\label{bc2}
{\Omega^{i}}_{\mu}={\Omega^{i}}_{\mu}+{\Upsilon^{i}}_{\mu}\,\quad {\Omega^{i}}_{\mu}\sim\left(\begin{array}{ccc}
 0 & 0 & -r/l \\ 
0 & -r_{+} r_{-}/r^3 & 0 \\ 
-r/l^2 & 0 & 0 
\end{array} \right)\,,\quad
{\Upsilon^{i}}_{\mu}\sim\left(\begin{array}{ccc}
 \mathcal O_{1} & \mathcal O_{2} & \mathcal O_{1} \\ 
\mathcal O_{2} & \mathcal O_{3} & \mathcal O_{2} \\ 
\mathcal O_{1} & \mathcal O_{2} & \mathcal O_{1} 
\end{array} \right)\,.
\ee
The above asymptotic behavior of triad fields are deriven from the Brown-Henneaux conditions \cite{BH} for asymptotically $AdS_3$ space-times
\be
{g}_{\mu\nu}={g}_{\mu\nu}+{G}_{\mu\nu}\,,\quad {g}_{\mu\nu}=\left(\begin{array}{ccc}
 r^2/l^2 & 0 & 0 \\ 
0 & -l^2/r^2 & 0 \\ 
0 & 0 & -r^2 
\end{array} \right)\,,\quad
{G}_{\mu\nu}\sim\left(\begin{array}{ccc}
 \mathcal O_{0} & \mathcal O_{3} & \mathcal O_{0} \\ 
\mathcal O_{4} & \mathcal O_{4} & \mathcal O_{3} \\ 
\mathcal O_{0} & \mathcal O_{3} & \mathcal O_{0} 
\end{array} \right)\,,
\ee
although it is not a dynamical variable in the Hamiltonian formalism. Including the black hole geometries in the asymptotic limit demands that the asymptotics (\ref{bc1}) and (\ref{bc2}) under PGT behave as   
\bea \label{pgtbc1}
\delta e^{i}\!\!\!&\sim&\!\!\!\left(\begin{array}{ccc}
 r/l-(r_{+}^2+r_{-}^2)/2 \,r l & 0 & 0 \\
 \\ 
0 & l/r+l (r_{+}^2+r_{-}^2)/2\, r^3& 0 \\ 
\\
-r_{+} r_{-}/r1 & 0 & r 
\end{array} \right)\,,\eea
and
\bea \label{pgtbc2}
\delta \Omega^{i}\!\!\!&\sim&\!\!\!\left(\begin{array}{ccc}
 0 & 0 & -r/l+(r_{+}^2+r_{-}^2)/2\, r l \\
 \\ 
0 & -r_{+} r_{-}/r^3 & 0 \\ 
\\
-r/l^2 & 0 & -r_{+}r_{-}/r l 
\end{array} \right)\,,
\eea
while for the vacuum configurations, these are $ \delta b^{i}=0,\delta \Omega^{i}=0 $. For the above boundary conditions, the solutions of the transformations (\ref{PGT}) are given by
\bea \label{diff} \xi^{0}\!\!\!&=&\!\!\!l\left[T+\frac12 \left(\frac{\prt^2 T}{\prt t^2}\right)\frac{l^4}{r^2}\right]+\mathcal O_4\,,\nn\\
\xi^{1}\!\!\!&=&\!\!\!-l\,\left(\frac{\prt T}{\prt t}\right)r+\mathcal O_{1}\,,\\
\xi^{2}\!\!\!&=&\!\!\!S-\frac12 \left(\frac{\prt^2 S}{\prt \varphi^2}\right)\frac{l^2}{r^2}+\mathcal O_4\,,\nn\eea
where the functions $T(t,\varphi)$ and $S(t,\varphi)$ satisfy the following conditions:
\be \frac{\prt T}{\prt \varphi}=l \frac{\prt S}{\prt t}\,,\qquad \frac{\prt S}{\prt \varphi}=l \frac{\prt T}{\prt t}\,. \ee

These relations lead to the periodic conditions
\be \label{pc1} \prt_{\pm}(T\mp S)=0\,,\ee
where $x^{\pm}=t/l\pm \varphi$ are light-cone coordinates. The commutation relation of the transformations (\ref{PGT}) is closed and produces a Lie algebra,
\be \label{pbpgt}[\delta'_0,\delta ''_0]=\delta'''_0(T''',S''')\,,\ee
and to lowest order we obtain
\bea \label{cr} 
T'''\!\!\!&=&\!\!\!T'\prt_2S''+S'\prt_2T''-T''\prt_2 S'-S''\prt_2 T'\,,\nn\\S'''\!\!\!&=&\!\!\!S'\prt_2S''+T'\prt_2T''-S''\prt_2 S'-T''\prt_2 T'\,.
\eea

The improved form of the gauge generator (\ref{GG}) is $\tilde{G}=G+K$, where the surface boundary term 
\be K=\oint df_{\alpha} \left(\xi^{0} \mathcal E^{\alpha}+\xi ^2 \mathcal M^{\alpha}\right)={\int_{0}}^{2\pi} d\varphi (l T \mathcal E^1+S \mathcal M^1)\,,\ee
depends only on the leading terms in $T$ and $S$ and not on all the gauge transformations in (\ref{diff}).
The non-trivial conserved charges at the boundary are the energy and angular momentum of the BTZ black hole. As mentioned before, these quantities are computed from (\ref{eam}) by choosing $\xi^{0}=1$ and $\xi^2=1$, and substituting the asymptotic values (\ref{bc1})-(\ref{pgtbc2}) in (\ref{eam}), we find the following expressions:
\bea \label{eam1}
\mathcal E^{\alpha}\!\!\!&=&\!\!\!2\, \varepsilon^{0\alpha\beta}\,\left[(\sigma+\alpha C) {\Omega^{0}}_{2}-\mu C(\frac32+\alpha (2\sigma+\alpha C)){e^{0}}_{2}-\frac{1}{\mu l}{\Omega^{2}}_{2}+\frac{1}{l}(\sigma+\alpha C){e^{2}}_{2}\right]{e^{0}}_{0}\,,\nn\\
\mathcal M^{\alpha}\!\!\!&=&\!\!\!- 2\,\varepsilon^{0\alpha\beta}\left[(\sigma+\alpha C) {\Omega^{2}}_{2}-\mu C(\frac32+\alpha (2\sigma+\alpha C)){e^{2}}_{2}-\frac{1}{\mu l}{\Omega^{0}}_{2}+\frac{1}{l}(\sigma+\alpha C){e^{0}}_{2}\right]{e^{2}}_{2}\,.
\eea

We apply the relations $\Omega=\omega+\alpha\,h$ and (\ref{he}) with additionally the prefactor $\frac{1}{8\pi G}$ which yield
\bea \label{eambtz}
E\!\!\!&=&\!\!\!\frac{1}{2G}\left[(\sigma+\alpha C) \frac{r_{+}^2+r_{-}^2}{2 l^2}+\frac{1}{\mu l}\,\frac{r_{+} r_{-}}{l^2}\right]\,,\nn\\ 
J\!\!\!&=&\!\!\!\frac{1}{2G}\left[(\sigma+\alpha C)\frac{r_{+} r_{-}}{l} +\frac{1}{\mu l}\,\frac{r_{+}^2+r_{-}^2}{2 l}\right]\,,
\eea
where $G$ is the positive 3D Newton constant. These values are exactly consistent with the relations in (4.41) of Ref. \cite{Skenderis:2009nt}, calculated for the BTZ black hole in the TMG case ($\alpha=0$) from holographic considerations. A relevant calculation of the conserved charges has been done in Ref.\cite{Tekin:2014jna} from the linearization of the equations of motion in the metric formalism.
\subsubsection{Asymptotic canonical algebra}
It has been shown that a suitably defined covariant poisson bracket (PB) algebra of the charge generators forms a centrally extended representation of the asymptotic symmetry algebra \cite{BH}. The PB of the gauge generators $\tilde{G}[\xi]$ is isomorphic to the Lie algebra of the asymptotic symmetries (\ref{pbpgt}), but in general, it has a centrally extended term. In fact for two sets of gauge generators, $\tilde{G}'\equiv \tilde{G}[T',S']$ and $\tilde{G}''\equiv \tilde{G}[T'',S'']$, the PB of the form $\{\tilde{G}'', \tilde{G}'\}$ is itself a differentiable generator. Since each differentiable generator can defined up to a constant phase-space functional $k$, so this bracket leads to 
\be \label{PB}\{\tilde{G}'', \tilde{G}'\}= \tilde{G}'''+k\,,\ee
where this central extended term known as the $central\,\, charge$ of the PB algebra.

In the canonical algebra when the constraints do not change under gauge transformations $\delta_0$, we can approximate the PB as $\{\tilde{G}'', \tilde{G}'\}= \delta'_0\tilde{G}''\approx \delta'_0 K''$. So getting together $\tilde{G}'''\approx K'''$ and (\ref{PB}) we have
\be \label{pc2} \delta'_0 K''\approx K'''+k\,.\ee

According to the periodicity conditions (\ref{pc1}), we can define $f^{\pm}=T\pm S$ and $U_{\pm}=(l \mathcal E^{1}\pm \mathcal M^1)/2$, so the following asymptotic transformations under $\delta'_0 K''$ are found,
\be \delta U_{\pm}=-f^{\pm} \prt_{\pm}U_{\pm}-2(\prt_{\pm} f^{\pm}) U_{\pm}+\left((\sigma +\alpha C){\mp}\frac{1}{\mu l}\right)\prt_{\pm}^3 f^{\pm}\,,\ee
where the last term is related to the second term of (\ref{pc2}) as
\be k=k_{-}+k_{+}=-\left((\sigma +\alpha C){+}\frac{1}{\mu l}\right)\int_{0}^{2\pi}d\varphi (\prt_{-}^3 f'^{-})f''^{-}-\left((\sigma +\alpha C){-}\frac{1}{\mu l}\right)\int_{0}^{2\pi}d\varphi (\prt_{+}^3 f'^{+})f''^{+},\ee
and $\prt_{\pm}=(l \prt_{t}\pm\prt_{\varphi})/2$ . We can define the Fourier modes as
\be L_{n}^{\pm}=-\tilde{G}[f^{\pm}=e^{i n x^{\pm}}]\,,\ee
and likewise the asymptotic generator is a linear composition as
\be \tilde{G}=-\sum^{+\infty}_{-\infty} (a_{n} L_{n}^{+}+\bar a_{+} L_{n}^{-})\,,\ee
where the periodic behavior of functions $f^{\pm}$ is defined in accordance with the condition (\ref{pc1}). 

The canonical algebra constructed from these Foureir modes takes the form of two Virasoro algebras as
\bea
\{L_{n}^{+},L_{m}^{+}\}\!\!\!&=&\!\!\!-i(n-m) L_{n+m}^{+}-i \frac{c_{L}}{12} n^3 \delta_{n+m,0}\,,\nn\\
\{L_{n}^{-},L_{m}^{-}\}\!\!\!&=&\!\!\!-i(n-m) L_{n+m}^{-}-i \frac{c_{R}}{12} n^3 \delta_{n+m,0}\,,\nn\\
\{L_{n}^{+},L_{m}^{-}\}\!\!\!&=&\!\!\!0\,,
\eea
where by `` $i\{,\!\}\rightarrow [,\!]$ " it takes the standard form such that $c_{L}$ and $c_{R}$ are left-right central charges
\be \label{cc1} c_{L}=\frac{3l}{2G}\left(\sigma+\alpha C-\frac{1}{\mu l}\right)\,,\qquad c_{R}=\frac{3l}{2G}\left(\sigma+\alpha C+\frac{1}{\mu l}\right)\,,\ee
which are exactly the values obtained in Ref.\cite{Bergshoeff:2014pca}.

\subsection{Thermodynamics}
The thermodynamical variables for the BTZ black hole such as the Hawking temperature and angular velocity of the event horizon $r_{+}$ are given by
\be \label{TV} T_{H}=\frac{1}{2\pi}\,\kappa\,=\frac{r_{+}}{2\pi l^2}\,(1-\frac{r_{-}^2}{r_{+}^2})\,,\qquad \Omega_{h}={\frac{1}{l}\,N_{\varphi}|}_{r=r_{+}}=\frac{r_{-}}{l\, r_{+}}\,,\ee
where $\kappa$ is the surface gravity in the ADM form
\be \kappa={\frac{1}{l}\sqrt{g^{rr}} N'|}_{r=r_{+}}\,.\ee

The values of the energy and angular momentum from (\ref{eambtz}) and thermodynamical parameters in (\ref{TV}) satisfy in the Smarr-like formula
\be \label{smf} E=T_{H} S_{BTZ}+\Omega_{h} J\,,\ee
and therefore the entropy of BTZ black hole for the MMG should be
\be \label{ent} S_{BTZ}=\frac{A_{H}}{4G}\left[(\sigma+\alpha C)+\frac{1}{\mu l} \frac{r_{-}}{r_{+}}\right]\,,\ee
where $A_{H}=2\pi r_{+}$ is the area of the event horizon. As seen, the result is consistent with Ref.\cite{BH} in the case of TMG when we set $\alpha=0$ in (\ref{ent}).

This value is also consistent with the Cardy formula from the holographic considerations. As shown in the previous section, the central charges of the dual conformal field theory are given by (\ref{cc1}). So from the Cardy formula 
\be \label{cardy} S=\frac{\pi^2}{3}(c_{L} T_{L}+c_{R} T_{R})\,,\ee
and that the left and right temperatures for the BTZ black hole are \cite{Maldacena:1998bw}
\be T_{L}\equiv\frac{r_{+}-r_{-}}{2\pi l}\quad\,,\quad T_{R}\equiv\frac{r_{+}+r_{-}}{2\pi l}\,,\ee
we obtain again the entropy (\ref{ent}). 

For this solution, the physical parameters given by relations (\ref{eambtz}), (\ref{TV}), and (\ref{ent}) satisfy a modified form of the differential first law of black hole thermodynamics\cite{Moussa:2003fc}: 
\be \label{mfl} dE=2\,T_{H} dS+\Omega_{h} dJ.\ee
In the calculation of the entropy in MMG, the same result is obtained by another approach in Ref.\cite{Setare:2015pva}.
\section{Conclusions}
In this paper, we have considered the asymptotic structure of the MMG in the canonical first-order formalism. Although this theory has an additive multiplier field $h$, it has the same degrees of freedom as TMG for $(1+\sigma \alpha)\neq 0$. From the field equations, it has been shown that, in order to have a torsion-free gravity, one can define a new $1$-form field $\Omega=\omega+\alpha\,h$. Then, we construct the canonical Hamiltonian of the MMG Lagrangian (\ref{LMMG2}) by defining the canonical momenta (\ref{cms}) for different dynamical fields.

Using the primary first-class constraints and appropriate Poincar\'{e} gauge transformations, we obtained the gauge generators (\ref{GG}) in according to the procedure in Ref.\cite{Castellani:1981us}. The asymptotic variation of generators under PGT gives some total derivatives. The contribution of $G_2$ vanishes, while the one for $G_1$ has a finite term after using the Stokes theorem. After all, we have found expressions for the conserved charges (\ref{eam}) and (\ref{eam2}) in the asymptotic region. 

We calculated the energy and angular momentum of the BTZ black hole in this formalism, the values of which are given by (\ref{eambtz}). These conserved charges are achieved by suitable asymptotic boundary conditions (\ref{bc1})-(\ref{pgtbc2}) for the canonical fields $e^{i}_{\mu}$ and $\Omega^{i}_{\mu}$. The PB of the improved generators $\tilde{G}=G+K$ has produced two versions of Virasoro algebra with two different central terms (\ref{cc1}), which is consistent with the asymptotic symmetry group of locally $AdS_3$ BTZ solution in Ref.\cite{BH}. One can easily see that the insertion of $\alpha=0$ in (\ref{cc1}) gives the values in the case of TMG.

Typically, the entropy of black holes in higher curvature gravitational theories are computed from the Wald formula. Since MMG does not have a metric Lagrangian, we found the entropy by using the Smarr formula (\ref{smf}). We also consider that the resultant entropy (\ref{ent}) accompanied by the energy and angular momentum satisfy in a modified first law of black hole thermodynamics (\ref{mfl}). To confirm this entropy to be truly consistent, we compare it with the expression obtained by the Cardy formula (\ref{cardy}) in the dual CFT. 
 
\section*{Acknowledgment}
I would like to thank  A. Ghodsi and M. Roshan for helpful comments. I am specially grateful to A. Ghodsi for reading the manuscript carefully.

\appendix 
\section{The algebra of constraints}
  The necessary and sufficient conditions for $G$ as a gauge generator are \cite{Castellani:1981us}
	 \be
	 G=primary,\qquad
	 \left\{G,H\right\}=primary,\qquad
	 \left\{G,any\,constraint\right\}=constraints.
	 \ee
The Hamiltonian equations yields the following constraints as (\ref{pc}) and (\ref{sc}):
\be \label{psc} {G|}_{\phi_{\rho}=0}=0\,,\qquad{\{G,H_{c}\}|}_{\phi_{\rho}=0}=0\,,\ee
where $\phi_{\rho}$'s are primary constraints and $H$ is the canonical Hamiltonian of the system in the gauge theory.
The PB algebra of these primary and secondary constraints (${\phi_{i}}^{\alpha}\,,{\Phi_{i}}^{\alpha}\,,{\psi_{i}}^{\alpha}\,,\mathcal H_{i}\,,\mathcal K_{i}\,,\mathcal T_{i}$) are given by 
\bea \label{ppba}
\{{\phi_{i}}^{\alpha},{\Phi_{j}}^{\beta}\}\!\!\!\!&=&\!\!\!\!\sigma \varepsilon^{0\alpha\beta} \eta_{ij} \delta\,,\quad \{ {\phi_{i}}^{\alpha},{\psi_{j}}^{\beta}\}=- \varepsilon^{0\alpha\beta} \eta_{ij} \delta\,,\quad \{{\Phi_{i}}^{\alpha},{\Phi_{j}}^{\beta}\}=-2 \mu^{-1} \varepsilon^{0\alpha\beta} \eta_{ij} \delta\,,\nn\\
\{{\psi_{i}}^{\alpha},\mathcal H_{j}\}\!\!\!\!&=&\!\!\!\!\varepsilon^{0\alpha\beta}(- \eta_{ij} {\prt}_{\beta}+2\varepsilon_{ijk} {h^{k}}_{\beta})\, \delta\,,\,\,
\{{\psi_{i}}^{\alpha},\mathcal K_{j}\}=-2\,\varepsilon^{0\alpha\beta}\varepsilon_{ijk} {e^{k}}_{\beta} \delta\,,\,\,
\{{\psi_{i}}^{\alpha},\mathcal T_{j}\}=-4\,\varepsilon^{0\alpha\beta}\varepsilon_{ijk} {e^{k}}_{\beta} \delta\,,\nn\\
\{{\phi_{i}}^{\alpha},\mathcal H_{j}\}\!\!\!\!&=&\!\!\!\!-2\,\Lambda_0 \varepsilon^{0\alpha\beta} \varepsilon_{ijk} {e^{k}}_{\beta} \delta\,,\quad \{ {\phi_{i}}^{\alpha},\mathcal K_{j}\}=2\,\sigma \varepsilon^{0\alpha\beta} \eta_{ij} {\prt}_{\beta} \delta-2\,\varepsilon^{0\alpha\beta}\varepsilon_{ijk}(\sigma {\omega^{k}}_{\beta}+\Lambda_0{e^{k}}_{\beta})\,\delta\,,\nn\\ 
\{{\phi_{i}}^{\alpha},\mathcal T_{j}\}\!\!\!&=&\!\!\!-2\, \varepsilon^{0\alpha\beta} \eta_{ij} {\prt}_{\beta} \delta+2\,\varepsilon^{0\alpha\beta}\varepsilon_{ijk}({\omega^{k}}_{\beta}-2{h^{k}}_{\beta})\,\delta\,,\\
\{{\Phi_{i}}^{\alpha},\mathcal H_{j}\}\!\!\!\!&=&\!\!\!\!2\,\sigma \varepsilon^{0\alpha\beta} \eta_{ij} {\prt}_{\beta} \delta-2\,\varepsilon^{0\alpha\beta}\varepsilon_{ijk}(\sigma {\omega^{k}}_{\beta}-{h^{k}}_{\beta})\,\delta\,,\nn\\
\{ {\Phi_{i}}^{\alpha},\mathcal K_{j}\}\!\!\!&=&\!\!\!2\,\mu^{-1} \varepsilon^{0\alpha\beta} \eta_{ij} {\prt}_{\beta} \delta+2\,\varepsilon^{0\alpha\beta}\varepsilon_{ijk}(\sigma -\mu^{-1})\,{\omega^{k}}_{\beta}\delta\,,\nn\\
\{ {\Phi_{i}}^{\alpha},\mathcal T_{j}\}\!\!\!\!&=&\!\!\!\!2\,\varepsilon^{0\alpha\beta}\varepsilon_{ijk} {e^{k}}_{\beta}\delta\,,\nn
\eea
where $\prt$ is the partial derivative and $\delta$ refers to $\{ {e^{i}}_{\mu}({\bf x}),{\pi_{j}}^{\nu}({\bf x}')\}={\delta^{i}}_{j}\,{\delta^{\nu}}_{\mu} \delta({\bf x}-{\bf x}')$.

\end{document}